# From Attention to Participation: Reviewing and Modelling Engagement with Computers

**Marc Miquel Ribé**
marcmiquel@gmail.com
Universitat Pompeu Fabra, Department of Communication, 138 Roc Boronat, 08018, Barcelona, Catalonia, Spain

**ABSTRACT**
Over the last decades, the Internet and mobile technology have consolidated the digital as a public sphere of life. Designers are asked to create engaging digital experiences. However, in some cases engagement is seen as a psychological state, while in others it emphasizes a participative vein. In this paper, I review and discuss both and propose a new definition to clarify the concept engagement with computers. Thus, engagement is a quality of an active connection between a user and a computing product – either a website or a mobile phone app. Studying it requires understanding a set of aspects like the user's affect, motivation and attention, as well as the product's design, content and composition. Finally, I propose explaining these concepts aligned with engagement and integrate them into a preliminary model to measure the manifestations.

**Keywords**: HCI theory, concepts and models; User Experience; Digital Media; Motivation; Engagement

## 1. Introduction

Understanding engagement has become the ultimate challenge for any designer or technology researcher, a sort of deep knowledge they all aim to. An engaging object is not just preferred over a similar one, but it will be more intense in any possible given use. Engagement means more. The term is employed in very different contexts, from games (E. A. Boyle, Connolly, Hainey, & Boyle, 2012; Cheung, Zimmermann, & Nagappan, 2014) to social networking sites (Freyne, Jacovi, Guy, & Geyer, 2009) and educational multimedia presentations (Jacques, 1995), among many others. In the web sphere, research has been particularly prolific while the industry has put analytics methodologies at the service of marketing objectives (Peterson & Carrabis, 2008). Engagement occurs in the highest complexity of virtual worlds, but also in the simplicity of a text-based communication. It has become a popular term synonymous of desirable.

Because of this, during the past years, empirical research has reached maturity and a great range of methods to study engagement and its dimensions have been detailed. The user has been analysed in its cognitive, emotional and behavioural dimensions, by means of both objective and subjective measures (Lalmas, O'Brien, & Yom-Tov, 2014). Likewise, the study of scenarios like multitasking (Lehmann, Lalmas, Dupret, & Baeza-Yates, 2013) or the use of multiple portable devices (Giang, Hoekstra-Atwood, & Donmez, 2014) have provided valuable insights on how people relate with technology. However, despite its soundness, empirical research has appeared dispersed and unable find a common ground for the studies. Paradoxically, although engagement gained momentum in empirical research, it remained vague at a conceptual level.

In fact, the broad use of the concept is at risk of overlapping with previous terms from the Human-Computer Interaction field. For instance, positive psychology Flow theory explains a mental state of a long and sustained use of an object with a focused attention, which is sometimes equated with engagement, but this is not necessarily the only way of engaging with objects. A more narrative-explorative use of the term engagement is explained by Activity Theory (Marsh & Nardi, 2014). Furthermore, in the past years, engagement has been used in Social Media websites such as online news to imply participation (Ksiazek, Peer, & Lessard, 2014; Liikkanen & Salovaara, 2015).

This participatory type of engagement has a long tradition in the field of Social Sciences, where civic engagement refers to the objective of involving citizens into participating in public affairs, or employee engagement focuses on worker performance. Even though with the emergence of Social Media this participatory type of engagement has become very popular, there is no model which explains how it occurs. The current framework for engagement with everyday websites is useful in order to explore the user experience (O'Brien & Toms, 2008), but does not include any concept dedicated to the intensity of interaction, namely the user's participation.

In this paper, I pursue the objective of defining engagement with computers and creating a preliminary conceptual model to encompass both attention and participation, based on the current, which I believe it can also be helpful to researchers, designers and users in reaching a common understanding. To study engagement, I propose and explain several essential aspects of the object (composition, design, logics and content) and of the user (emotion, motivation, understanding and attention). The main contribution of this paper is a theoretical discussion leading to a preliminary model, which ultimately bridges theoretical concepts with current empirical research. In general, I see this as one more step towards a better understanding of how people engage with technology. This paper is organized as follows:

In Section 1.1, I review the main definitions and background of the different uses of the concept ‘engagement’ in Social Sciences and Human-Computer Interaction. Then, in Section 2, I propose a definition of engagement of computers and a discussion of each of the aspects which influence it. I integrate such aspects into the model. Finally, in Section 3 I conclude with a discussion.

### 1.1 Previous definitions and applications

In this section, I review the definitions and applications of engagement to better understand how the concept has been used and what challenges it may involve.

By definition, to *engage in* is "to attract and hold fast", while *to be engaged* stands for, among other possible meanings, "occupying the attention of someone", either with an activity or with a commitment[1]. From these multiple meanings, two separate streams of research on engagement arise, with a growing cross-fertilization between them: one in the interdisciplinary field of Human-Computer Interaction (HCI) which usually approaches engagement focusing on the psychological aspects of a person performing an activity with technology; the other in the broad field of Social Sciences, which remarks the person's commitment and social actions such as participation.

#### 1.1.1 Human-Computer Interaction tradition

The concept engagement was first employed at the beginning of the 1990s to characterize the user's psychological state while interacting with all kinds of technological interfaces. Therefore, it covered meanings similar to being attentive and absorbed while enjoying technology. First, Laurel (1991) studied software interfaces and referred to engagement as the feeling of being in direct manipulation with a physical object. Laurel (1991) considered that when a system is working properly, the user entails "sustained belief" that it will respond as if it is alive, even bringing "playfulness". Further on, in the context of educative technology, Jacques et al. (1995) referred to engagement as the effect of a system which ultimately attracts the user's attention by arousing his emotions. For Webster and Ahuja (2006), engagement with a website was similar to a flow state of mind in which the user enjoys a very focused attention, and its satisfaction could trigger a future intention to return to the website. Engagement was considered mainly an emotional or attentional component, but with a sense of amusement.

As technology evolved and computing products were designed for many more objectives, the term engagement incorporated "user" next to it. For instance, in video games, user engagement was considered a prior phase to immersion and presence (Brown & Cairns, 2004), two states in which the player abandons himself in a virtual world and identifies himself with the character. Still in video games, it was correlated with entertainment and it was very influenced by usability (E. A. Boyle et al., 2012; C. M. Karat, Karat, Vergo, & Pinhanez, 2002). Later, van Vugt et al. (2007) analysed engagement with virtual reality by measuring it as a concept between involvement and distance. In other distant fields like the creation and use of information systems, user engagement also comprised a sense of involvement (Hwang & Thorn, 1998; Kappelman & McLean, 1992). All in all, the term comprised different psychological attributes depending on the context and application, and it overlapped with other concepts in the same field of Human-Computer Interaction.

[1] http://www.dictionary.com/browse/engage

One of the overlapping concepts is the term user experience (UX), which appeared several years after engagement, to cover the emotional and exciting side of technology. The advantage of UX over previous concepts is that it allowed introducing a discourse which was not centred on efficiency (like usability). By taking into account the user psychological state, and also by emphasizing positive emotional outcomes such as joy, fun and pride (Hassenzahl & Tractinsky, 2006), UX found its place and dominated the field in the development of services, products and computers – both in the academia and especially in the web industry. The original sense of playfulness initially explained by engagement was then better covered and generally assumed to belong to UX. It became the popular term and the general catch-all term to refer to user needs, feelings, thoughts, expectations in order to improve the design process (Hassenzahl & Tractinsky, 2006).

Engagement needed to be redefined in order to avoid repeating the same UX debates in a parallel research line. A possible solution was given by O'Brien and Toms (2008) whose strategy was to define user engagement as "a quality of the user experience" (p. 949). By embedding engagement into the newer, more popular and studied concept of UX, O'Brien and Toms (2008) would limit the concept to a range of positive experiences. The above-mentioned authors developed a framework for the web research where user engagement is characterized by "challenge, aesthetic and sensory appeal, feedback, novelty, interactivity, perceived control and time, awareness, motivation, interest, and affect". This extensive list of attributes was very common to the UX studies - e.g., aesthetic appeal (Lavie & Tractinsky, 2004) and emotion (Forlizzi & Battarbee, 2004) -, and it would explain engagement modelled as a process, with a "point of engagement", an "engagement period", a "disengagement moment" and maybe a "re-engagement".

Even though the use of attributes could explain how a time-based process develops, I consider it presents several problems. My critique to this perspective is three-fold:

- First of all, the framework has varied along the years and the authors included usability as secondary when applying the framework to news portal (O'Brien, 2011), while it disappeared in later versions to include trust (Lalmas et al., 2014). It is necessary to clarify which are the essential attributes to explain engagement and which attributes are instead secondary to better understand specific scenarios.

- Second, the UX perspective of engagement solely considers the user, thus the attributes are often rewritten from this point of view even though they do not emanate from it. Usability becomes 'perceived usability' and aesthetics 'aesthetic and sensory appeal'. This relegates the object in a passive secondary plan. Hence, certain aspects of the object like 'content' or 'meaning' cannot be incorporated in the framework.

- Third and most importantly, possibly due to the embedding of engagement into user experience, there is no attribute related to the external dimensions of engagement such as user behaviour (interaction, or participation). Nevertheless, empirical research based on this framework ends up measuring user behaviour by using metrics and data analysis techniques (Attfield, Kazai, & Lalmas, 2011; Lalmas et al., 2014). If the intensity of the user behaviour is considered engagement, then the relationship with its causing factors should be explored. In other words, for a more comprehensive model of engagement, a participatory type should be explained.

### 1.1.2 Social sciences tradition

When engagement is applied to Social Sciences, it emphasizes the participation and a sense of social relatedness. Examples are varied from all areas of public life. For instance, in civic or political engagement (Ball, 2005), engagement implies an orientation or predisposition towards action. To engage citizens means helping them become members of the political process through discussions and debates which influence them. Any kind of community engagement refers to the way an individual integrates into a group, whether if it is a public government, an education system or a research group (Ahmed & Palermo, 2010). Engagement is desirable in order to improve social dynamics, give value to the relationships and achieve their group goals.

In addition, by engagement is also meant an individual process where the individual progresses in a specific activity or environment. In the education field, it is connected to intensity of behaviour and emotional involvement during the task (Appleton, Christenson, Kim, & Reschly, 2006). In a work environment, employee engagement is seen in terms of the relationship with the organization, of the commitment with group values, and it is aimed at improving group performance (Reeves & Read, 2009). Likewise, sport engagement is more focused on the path of achieving autonomy and improving the quality of its practice (Alvarez, Balaguer, Castillo, & Duda, 2009).

All kind of groups and individuals are interested in having engaged people, whether these people assume their activity consciously or the purpose is not publicized and goes unnoticed. This is especially interesting for all the fields related to business. In marketing, brand engagement refers to the relationship of a customer with the image of a product or company, encompassing aspects from the regularity of use, involvement, or even recommendation to others (Arcas, 2014; McWilliams, 2013). Similarly, customer engagement discusses how users co-create value around a company, purchases and interactions (Brodie, Hollebeek, Juric, & Ilic, 2011).

Until recently, any activity would include the term engagement and associate it to participatory values, while technology use would refer to engagement as a matter of attention and emotion. However, the advent of Social Media and all the new technological and portable devices has led to a wide sense of the term engagement which overpasses frameworks and past definitions like the one from O'Brien and Toms (2008), which is narrowed to attributes from the cognitive and emotional dimensions of the user without including the behavioural dimension.

Especially in the web sphere, examples of new forms of engagement based on this social sense are abundant. For Peterson and Carrabis (2008), visitor engagement in websites implied reaching some objectives throughout the measurement of user behaviour with metrics (e.g. number of pages visited, loyalty or recency). In social networking sites, engagement is mainly considered and measured in terms of social interaction - i.e. number of votes, comments or shares - (Smith & Gallicano, 2015), while in content repositories like Wikipedia, the editor engagement is linked to the editing activity in articles and in policies (Halfaker, Geiger, Morgan, & Riedl, 2013a), as their success is totally dependent on it.

### 1.1.3 Summary and challenges

In summary, research on engagement has been conducted both on individuals and on groups (in the latter case in organizations with collective goals) interacting with all kind of objects. Engagement happens 'to be everywhere' because it is implied in the sense of relating to something. When applied to the current technology, I remark the following shared conclusions about engagement from both Human-Computer Interaction and Social Sciences traditions:

- Engagement is an objective of the researcher or the designer, who all have an expectation set on the user to act in a particular way. Therefore, the measurement of metrics (Peterson & Carrabis, 2008) confirm an object is properly designed for its goals.

- Engagement is multidimensional in the emotional, cognitive and behavioural aspects of the user (Attfield et al., 2011; Lalmas et al., 2014), and also takes into account the design aspects of the object such as usability.

- Engagement is considered positive with no clear absolute value. It has a positive sense which emphasizes the positive aspects between both object and user interactions (Lehmann, Lalmas, Yom-Tov, & Dupret, 2012).

As seen, the obstacles for obtaining a universal definition of engagement reside in the slightly different uses of the concept in non-related fields, in the technological advances and their socialization, in addition to the interferences from non-academic uses of the word. The most important challenge is that current models do not explain the participatory type of engagement. In order to conciliate this weakness, I attempt to provide a clear definition to study engagement. This will be the object of the next section.

## 2. Engagement with Computers

In this section I propose a new definition for the concept engagement with computers and I discuss the aspects that influence it.
Consistently with the aforementioned conclusions from past section, I define engagement as ***the quality which guarantees that the connection between a user and a computing product remains active.*** In doing so, the concept of engagement becomes inclusive of the previous uses in both tradition Human-Computer Interaction and Social Sciences, and fits best with the available evidence from research. Engagement exists as long as the connection is alive.

The computing product - term used by {Fogg:2003et} to refer to objects from websites to apps - necessitates the user's responses (the minimal response being attention). The user behaviour can be either the user passively absorbed or participating frenetically, but in both cases, it guarantees the connection remains active. This way, the user participation becomes one specific manifestation of the connection.

An engaging computing product or object is desirable or alluring, because it ensures the user's attention and it keeps the connection alive. Checking the e-mail, updating a profile in a social networking site or browsing the Internet in the search for a particular piece of information can imply connections at different levels of engagement intensity. Thus, the expressions of engagement are the outer manifestations of the connection, in other words, the interactions between the user and the computing product that keep the connection active. Engagement is a concept to understand such connection in its multiple configurations.

Each connection may manifest itself in a different way (longer or shorter duration, and more or less interaction). These manifestations are measurable and can be explained by studying each of the two parts of the connection. Hence, engagement needs to be holistic and embrace complexity, as all user and computing products aspects are interrelated and may influence one another. For a full understanding of the engagement quality, one has to consider both user and computing product aspects. These will be discussed in the following sections.

### 2.1 Study of the Connection

**User, object and agency.** Drawing upon this definition, I view the connection between user and computing product as the unit of analysis, where user and computing product (or simply object) are equally important. This is in sharp contrast to the user-centred paradigm prevailing in Human-Computer Interaction, which considers computing products as passive tools. The first computing products designed for massive use had a practical goal. For instance, the spreadsheets allowed companies or families perform accountability calculations in an easier way. Later on, computers enabled the creation of objects simulating a place where users could engage in activities; this was called *media* or *medium* (Laurel, 1991).

Nowadays, websites and software create digital spaces in which the user learns, plays, competes or communicates with others. And most remarkably, the last frontier in computing products is their capacity to perform communicative actions, which makes them convert into *social actors* - for instance, a personal assistant which can figuratively encourage users to achieve goals or change habits in their daily life (Fogg, 2003). This is why in the advent of a more sophisticated artificial intelligence, in order to understand the connection between a user and a object, engagement should not be exclusively centred on the user's perceptions, needs or behaviours, but it should also consider the object.

Computing products can be programmed to constantly change in their design and content to attract and maintain interaction with the user, and their behaviour can be totally unexpected as if they were beings (Suchman, 2007). Yet, they cannot respond to the notion of 'agent', or the "one who initiates the action" (Laurel, 1991, p. 4). They are designed with an active purpose, but when it comes to establishing a new connection, they are conditioned by a user's previous acceptance (e.g. a smartphone is able to receive app notifications but only if the user turns it on). One can say that computing products have no consciousness, but their intentionality is delegated by designers to its content (and meaning), design (both aesthetics and interaction), and behaviour (sustained by the logics of their algorithms). From the designer's perspective, the user has to confirm the quality of his creation with its use (i.e. engagement is a measure of success).

**Object composition.** Every computing product can be composed of smaller objects sometimes to provide a functionality or a new piece of information. Therefore, each inner object has at least one action to engage a user into a pattern of interaction and influence behaviour. The way computing products are composed is the key for simple and complex websites, video games and all kinds of computing products. Any connection with a compound object can lead to several sequential connections with different inner objects. This is known as multitasking, and it can happen either with multiple objects or with compound objects.

A social networking website is a clear example of computing product that is a compound object. This often includes inner objects such as a synchronous communication channel (i.e. "chats"), photographs and news all at once. Banhawi, Ali, & Judi (2012) analysed the use of the social networking site Facebook and found that the novel content, appearing constantly drives people to be eager to see more. The users' preferred activity was writing to other users' personal spaces, followed by watching photographs, status updates, social investigation and content surfing. Users engage with Facebook in unlimited combinations with inner objects which disappear or are substituted. Likewise, this can also happen in a context with multiple objects, in which the user has to respond to notifications from a social networking, the e-mail, an opened document, the phone and a control panel with a connected smart home (see Figure 1).

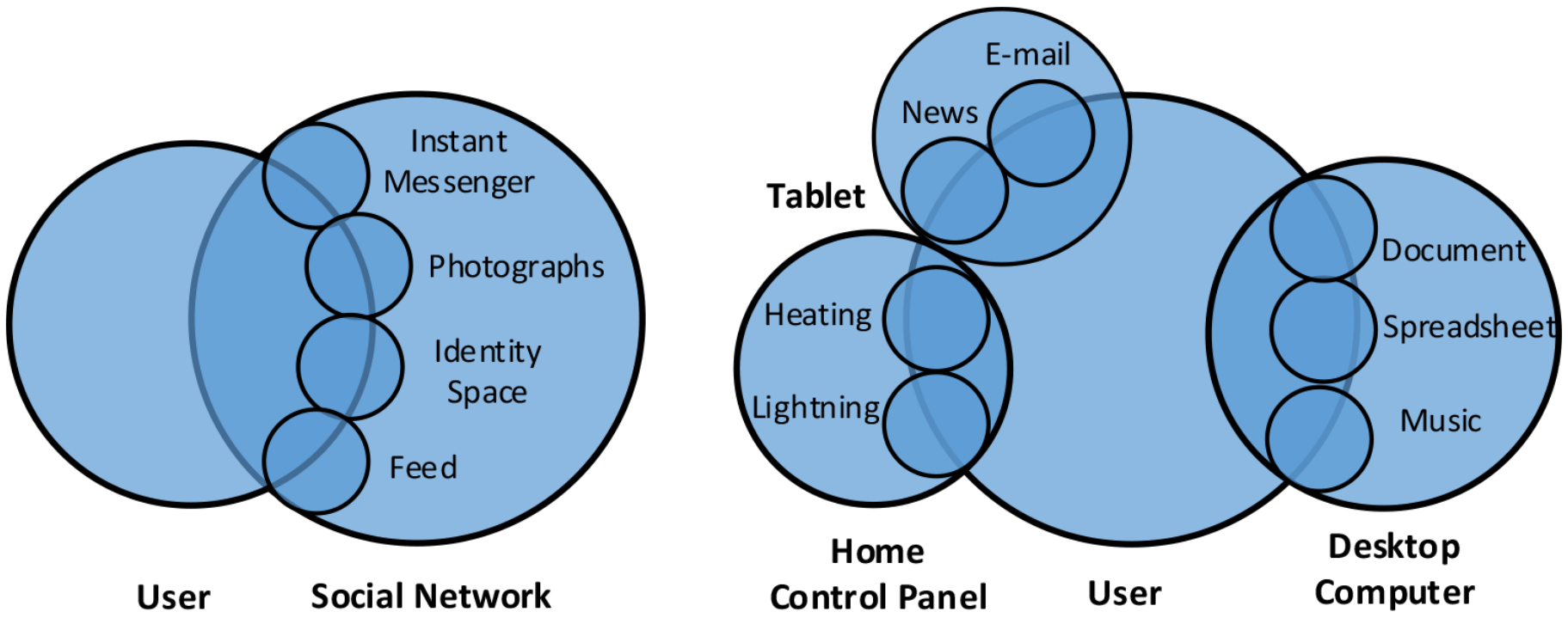

***Figure 1. Two scenarios of engagement with multiple objects. One with several inner objects from a Social Networking Site and another using different electronic devices***

**Aspects of engagement.** When developing any engagement study, it is mandatory to define the two parts (user and object or objects) and the precise context in which the connection takes place. The holistic view of engagement implies that all elements must be taken into account for their interrelations. For example, when a group of users connects to a single object, this is often called community engagement. Likewise, a single user can engage with multiple independent objects in order to reach a specific goal, to understand a story or simply for the sake of entertainment. Hence, the study of engagement can get beyond the limits of a single object and include multiple objects in the same scenario. As a result, focusing on only one object (e.g. a website) without considering the context of use (i.e. the rest of objects) would lead to wrong conclusions (Lehmann et al., 2013).

Once the two parts are specified, it is necessary to understand which are the inner aspects which drive them to constitute in a temporary relationship and maintain it. On the wake of O'Brien and Toms (2008) I aim to appeal at several aspects of the user and the object to explain the reason why the user and object stay connected. As far as the user is concerned, I propose ***emotion, motivation, cognition* and *attention***, as related concepts. When it comes to the object, I take into consideration ***composition, design, logics and content*** (Figure 2). The connection between both parts will generate a fluent dialogue, which is an aspect dependent on both parts. Each of these aspects will be developed and explained in the following sections, in order to set a proper setting, hypothesis, and the variables to perform any experiment.

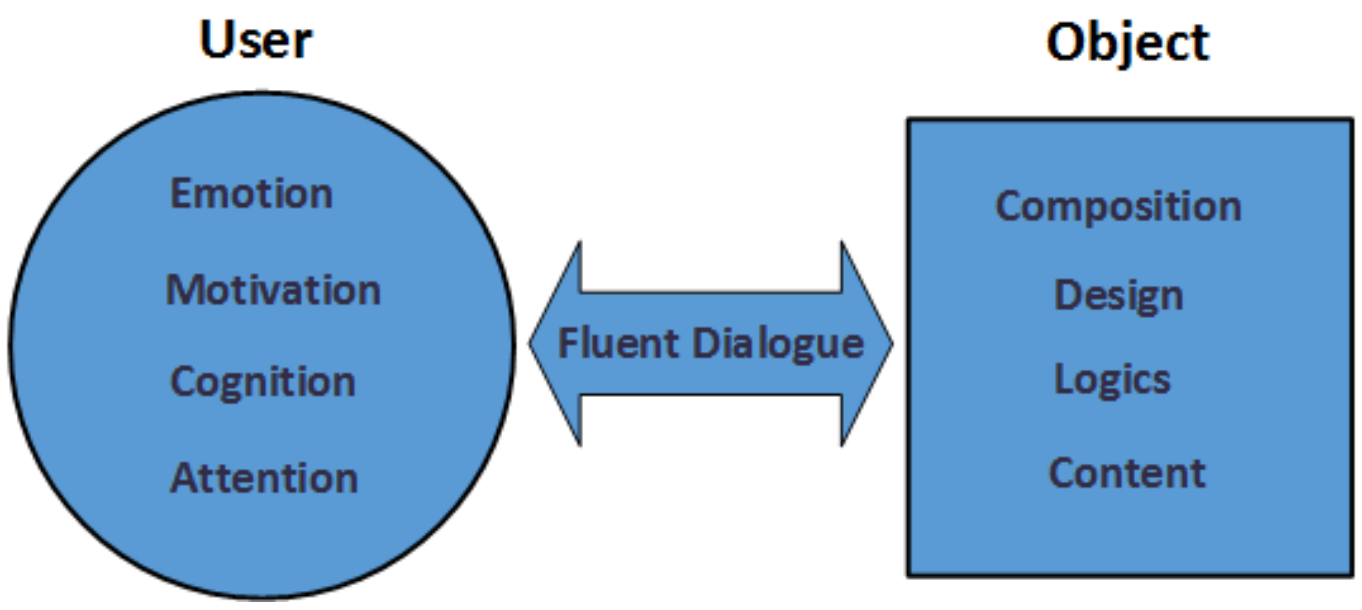

***Figure 2. Main aspects of the user and of the object influencing engagement.***

## 2.2 User's emotion and motivation

User's *agency* or *drive* to act has been widely explained by the user's emotional and cognitive dimensions. Concepts such as emotion and motivation are fundamental in understanding why a user gets involved in an activity with a object. In psychology, an emotion is seen as a set of internal processes of self-maintenance and self-regulation (Markus & Kitayama, 1991). The introduction of this concept in the study of technology use has contributed to understanding the centrality of emotion in the user's experience. Strong emotions and pleasure alter our perception of products (Forlizzi & Battarbee, 2004). User's positive emotions assure the connection is maintained and guarantee user's satisfaction at the end of the task; in the same way as user's good performance predicts a future intention to return (Chung & Tan, 2004; Webster & Ahuja, 2006).

Motivation is a complex construct linked to both emotion and cognition. Recent studies appealed to motivation to study the depth of engagement (Ainley, 2006; Bouvier, Lavoue, & Sehaba, 2015; Chapman & Selvarajah, 1999; T. de Vreede et al., 2013; O'Brien & Toms, 2008). Motivation is the key factor for users to initiate, persist in or resume an action. It is related to energy, direction, persistence, all aspects from activation to goal reaching (Deci & Ryan, 2012; Ryan & Deci, 2000). Every connection with an object has a motivation behind, whether it is random and unique access to a website, or routine and regular use of a phone App. The principle 'The more positive the experience, the more driving force will the object have' does not always apply. As a matter of fact, some experiences can be unpleasant or arouse negative emotions in the user and still motivate the user to engage with the object.

Since motivation is central to the user's behaviour, understanding it is at the very basis of understanding engagement. In other words, the study of motivation allows the researcher to explain how to make connections last longer or be more intense in terms of interaction, namely the specific design changes he would implement. There is a great variety of models of motivation; it is not a unitary phenomenon. For instance Self-Determination Theory relates motivation to psychological needs, such as relatedness, competence and autonomy (Ryan & Deci, 2000). The same theory proposed the distinction and generalization of motives into intrinsic and extrinsic, according to the user's locus of control towards action. Intrinsic motivation is independent from any valuation and is induced by the inherent satisfaction derived from performing an activity. On the contrary, extrinsic motivation is triggered by activities which imply an outcome of any kind, either a reward or ego involvement.

Concerning immersive experiences, it has been argued that an intrinsic motivation can easily lead to focused states of attention. Theories like Cognitive Evaluation Theory (Ryan & Deci, 2000) or Flow Theory (Csikszentmihalyi, 1991; Nakamura & Csikszentmihalyi, 2009) have explored which factors could facilitate an intrinsic motivation, focusing on user's competence and autonomy. Flow is achieved in an activity with challenges of all kind (mental or physical) but which does not exceed the user's existing skills, in such a way that the user is in control of the situation, never bored but neither anxious, between control and arousal. The user is capable of dealing with every challenge according to his skills, while new challenges appear continuously. With this mind-set, a user performs a task driven by intrinsic motivation, and he does so with such a joy and intense concentration that he loses reflective self-consciousness and sense of time (Nakamura & Csikszentmihalyi, 2009). This is a so-called "optimal experience" because the user does the best performance and the sense of absorption in the activity is complete - a loop in experience.

In fact, intrinsic motivation can also be a source of very joyful experiences (Chapman & Selvarajah, 1999) and its related flow state can be very beneficial to keep a user engaged and therefore a connection active. However, active connections can also exist in contexts that do not provide the suitable challenge-control structure necessary for Flow to happen (O'Brien & Toms, 2008; Webster & Ho, 1997): they can be fostered by extrinsic motives such as social or physical rewards. The importance of motivation lies in that it sets the direction for user's action and the reason behind it, which depending on the degree of motivation may act with more or less intensity. In consequence, a strong motivation will lead to long and sustained object use or, in other words, to an intense interaction. However, aspects regarding the object design, content and purpose can be as determinant as motivation on how the interaction unfolds in a connection.

### 2.3 Object's Design, Content and Logics

**Facilitating Flow and Zone.** The most significant difference between physical objects and objects is that the latter can be designed up to its minimal details. Design implies both aesthetics and functioning of the object. Changes in object design can be tailored to respond to the different kinds of motivation and improve engagement by providing interaction. Ever since their appearance, video games have been considered the closest expression to a complete digital and active reality. Przybylski, Rigby and Ryan (2010) applied the Self-Determination Theory to videogames and found out that they induced a feeling of well-being into players due to their addressing and fulfilling basic psychological needs of the user, such as competence, autonomy, and relatedness.

Motivation can be reinforced by interaction and design, and therefore both aspects contribute to maintaining the connection. One example can be found in Cheung, Zimmermann and Nagappan (2014), who evaluated the impact of different video game design elements by means of self-reported comments. They advocated the idea that design was crucial for engagement (especially during the first hour) and that it influenced how players perceived the rest of the game. Namely, according to the abovementioned authors, "the first hour must provide the right balance of challenge and skill to put players on the right track to enter a flow state" (p. 59). In this first hour, the player learns the control keys, the mechanics and the consistency of the scenario, which allows him to progress and gain control at the same time, satisfying his motivation. Among the players' comments collected by the study, several users were asking for trainings to be provided at the beginning of the game in order to avoid frustration. Cheung et al. (2014) concluded that it was the rapid figuring out of how to control interaction (clarity in the interaction controls), a curve of challenges as well as allowing the user to set further goals which kept motivation and interaction stimulated.

In a slightly different environment, Schüll (2012) studied a scenario involving no challenge, namely the games of chance. Schüll (2012) noted that the use of videogambling machines in Casinos induce a psychological state called 'Zone'. This state was very comparable to Flow in terms of absorbed attention, but instead of stimulating activity and high performance, the player remains in an idle and desubjectified position for hours. Paradoxically, in the zone the player seeks and feels a sense of control while actually being out of control (Schüll, 2012). Schüll attributes this state to the way the design conduces the player triggering extrinsic motivations (with reward structure). The odds of winning are low and even the frequency depends on optimized algorithms aimed to trap specific player preferences and styles. In addition, the interface with numerous buttons creates a false control sensation and every option is disposed to keep the zone going. When the player enters the game, he delegates the action to the machine.

In both experiences of Flow and the Zone, interaction design and motivation are crucial to reinforcing the connection, which remains active in a loop structured experience. The two cases present some structural differences and similarities (see **Error! Reference source not found.**). The main difference consists in the necessary challenge and sense of progress, which in well-designed videogames or in any other object may lead to flow while the user is trying to accomplish a goal. The need to be in control of the situation is a requirement for Flow, and it is based on a cause-effect sensation in each user action. The same does not occur in videogambling machines mainly because their game goals are set externally to user's will and are not controllable. Flow implies an emotional self-reward and intrinsic motivation, while the Zone is provoked by the videogambling rewards which entail monetary prizes (extrinsic motivation). Instead, the commonality they share is a sense of rapid feedback and a continuity. The feedback provided by the videogambling design is a key factor in maintaining the player's desire to continue. The fact of being shown the next gambling round triggers a passive acceptance in the player.

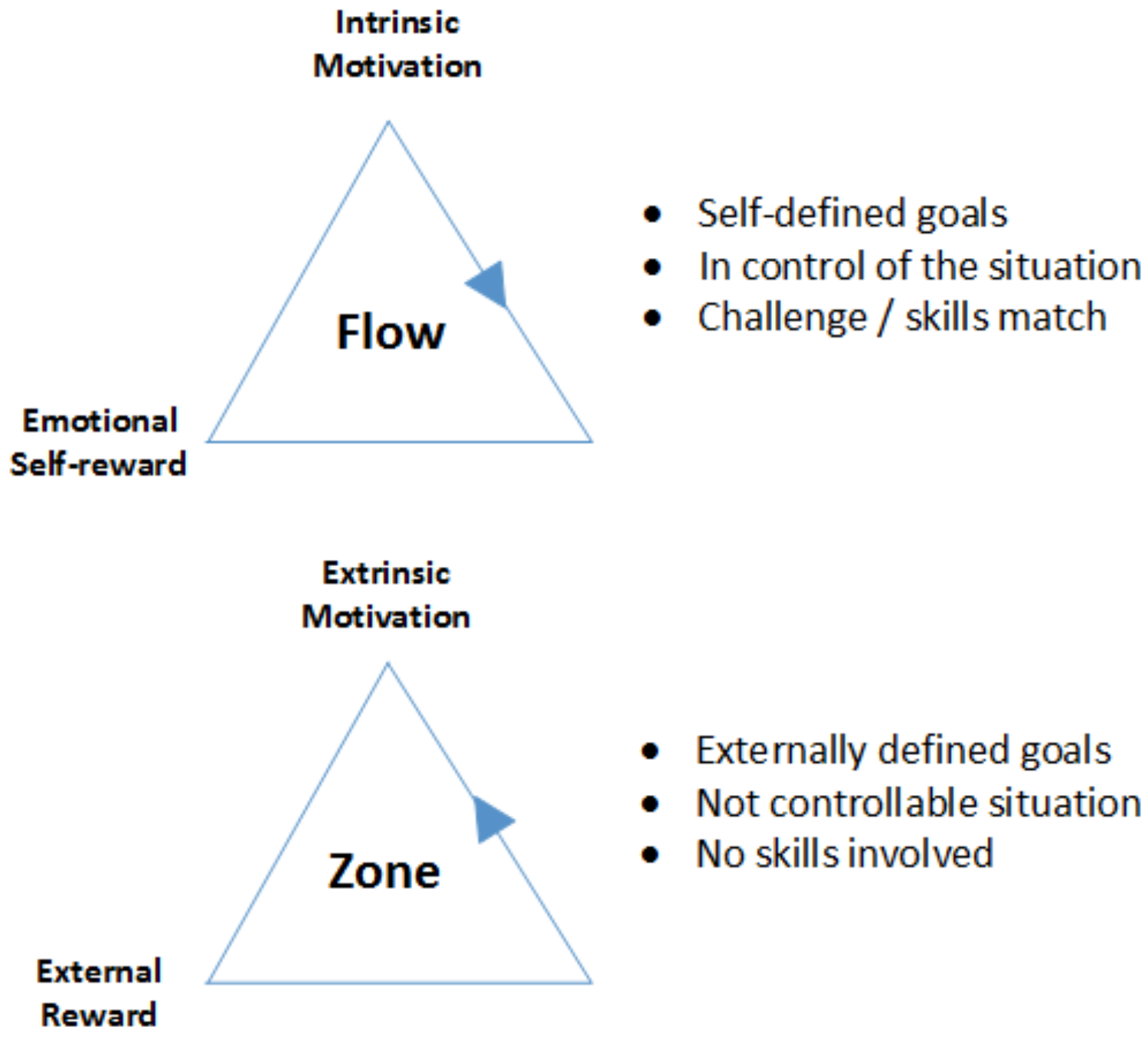

***Figure 3. Flow and Zone motivation loops and their structural characteristics.***

As already stated, in both cases design is the counterpart for motivation. When the user's skills and decisions play a determining role in the interaction, the continuity is totally *user-directed*, when instead this is not the case, the continuity is totally *object-directed*. As long as object's interaction design presents continuity and reinforcement for user's motivation, the motivation type can become secondary. In other words, the object, in order to feed the user's needs and motivation, can present feedback and affordances.

Flow and the zone are two clear and delimited types of immersive experience with an emotional and motivational loop structure, but certain sophisticated objects can produce similar effects. Mauri et al. (2011) investigated the psychophysiological effects of Facebook to find out why it is so successful. They noticed that the measures described a core state (between valence and arousal) very similar to Flow, but in an environment of no challenge. In such a social networking site, the positive affect is associated with a recreational activity which addresses the social needs of the user by presenting multiple inner objects related to the user (Mauri, Cipresso, Balgera, Villamira, & Riva, 2011).

All in all, user's motivation is central to engagement, but no less than the way the object design anticipates the interaction. In fact, a user can remain engaged with an object, switching between inner objects, as long as there is a motive for interaction (Marsh & Nardi, 2014). In this sense, some objects may be designed with strategies using a rich variety of characteristics and functionalities. When studying engagement, it may be interesting to ask: what contributes more to connection continuity, the user's motivation, or the interaction provided by the object's design? In some cases, it is the object that is more influent in keeping the connection going, while in other cases it is the user's motivation.

**Object design strategies for continuity.** Engaging with computing products is very similar to engaging in the physical world with people, processes, places or groups. However, the objects can count on strategies to encourage continuity by anticipating steps totally tailored to the user's motivations. These strategies can use content and meaning (the "what"), but also different design components and available actions (the "how").

In fact, **content** is the object's property which triggers the user's interest, a kind of intrinsic motivation with positive emotional valence (Ainley, 2006). In consequence, users can possibly be engaged because a specific content is interesting to them. As an example, in an online news website the specific content was a key factor engaging users in reading (Arapakis, Lalmas, Cambazoglu, Marcos, & Jose, 2014). The higher the users' interest, the more comments they posted, in parallel, the more enjoyment they draw from watching the video, the higher the possibilities to take an active role and comment. In a similar manner, de Vreede et al. (2013) considered that engagement in a crowdsourcing community was determined by how it enabled developing personal topic interest - stimulating the user to go from a passive user to an active contributor.

Other content strategies to maintain the user emotionally aroused aim at providing novelty or structuring information in the form of a story. Laurel (1991) studied the different canons of drama theory and showed that meaning could be a driver of interest in keeping attention high. The different phases of a linear story raise or lower the emotional arousal the same way in a narrative video-game as in a theatre play. The interaction between the elements of the story, if well written, cause in the user excitement and interest to see what comes next. In a object, these elements can be combined and varied according to inputs, while the experience can be personalized to keep the motivation high. This is the case of social networking sites: the content continuity is provided in a central channel of information (feed), while the social continuity is ensured by means of a synchronous communication channel (chat). In fact, putting the accent on computer-mediated communication in order to convert websites in social and foster engagement has been a common strategy.

In addition, technology can be used for persuasive purposes such as increasing engagement (Fogg, 2003). Sophisticated **logics** (algorithms) allow objects to perform actions humans would not be able to. For instance, objects can be persistent in presenting actions repeatedly and in an impersonal way (e.g. a software sending e-mail to customers informing them of an unfinished purchase and what they left in the basket). To this same purpose, objects can use rich design components based on video and sound. Not to mention their easiness of transport which grants ubiquity.

In general, the more technology evolved, the more engagement has become critically dependent on design aspects such as rapid feedback. I believe that with the development of artificial intelligence technologies and the abundance of data, designers' efforts will focus more on personalization, in order to achieve a greater symbiosis between the user and the objects. An example of this is the *filter bubble* algorithm used in web searches and social networking sites. This strategy exclusively provides results or information tailored according to previous results, avoiding cognitive dissonance and therefore reinforcing the user's point of view and expectations (Pariser, 2011).

### 2.4 Cognition, Usability and the Fluent Dialogue

Properly designed objects can entice interaction by providing new goals to keep the user motivated. In activities where the user's creativity is stimulated, motivation alone can be sufficient. However, besides being motivated, the user also needs to understand how to proceed. It is only after a repeated use of an object that the user internalizes actions and achieves autonomy to perform the activity with little effort or conscious thought (Marsh & Nardi, 2014). And even though most of the times the user can learn and become tech-savvy, the object is also expected to support the user by means of an understandable, self-explanatory design.

Having a fluent dialogue depends both on the user and on the object. It can be looked at in terms of a trade-off between the user's skills and cognition and the object's design with its affordances for interaction. While the user must identify the next step, and understand how to reach it, the object must provide clear affordance and feedback in order to facilitate the user in doing so. One of the object's usability goals is avoiding user disorientation by offering clear affordances on any possible further action. Any sort of feedback is useful to inform on the user's progress in attaining a specific goal.

Usability is the design property responsible for providing feedback, visual cues and information in order to facilitate the performance of the user and make it satisfying and memorable (Nielsen, 1999). In early studies feedback quality and speed were considered related to usability. For Laurel (1991), feedback was a necessary part in order to sense a "direct manipulation" with the interface, because it reinforced the interaction with immediate response.

On the contrary, lack of feedback leads to frustration, because it leaves the user with no indication on how to continue the interaction, and can be as detrimental for the connection as the lack of interest or of motivation. This is why usability is a central aspect of the object, and it has sometimes been taken for granted in relation to engagement in previous models (Lalmas et al., 2014; O'Brien & Toms, 2008).

Depending on the type of object, there may be varied design components available for interaction (audio, visual, touch, space, etcetera.). Fluent dialogue may exist in the different design components. For instance, a game can imply a 3D immersive experience with a whole range of audiovisual features; an instant messenger may only involve text and few pictographic images. In some cases, these components may even allow the user to modify the object (e.g. comments in a website or uploading a video) or communicate with other users.

If an object encompasses several inner objects, the design should consider the overall perception of these objects in order to avoid confusing the user. A fluent dialogue between a user and a compound object can take place with multiple channels and inner objects at the same time, in a similar way to multimodal communication (Klein, 2015; Norris, 2004). This could be the scenario of home automation, in which temperature, lighting and music are controlled coordinatively. For instance, song selection could be manipulated using a screen interface, while temperature change could be simultaneously activated by voice. Very importantly, in order to attain a fluent dialogue and keep connections active the diversity of components does not have to exceed the user's cognitive abilities.

## 2.5 The Connection is Reciprocity

Previously, I assumed that for a connection to be active there must be reciprocity between the computing product and the user. While the object is able to create and manage multiple connections with different users independently, the user can *only* respond to one connection to the exclusion of others - giving it his attention in a precise situation or in repeated moments along time. Attention is the cognitive process of selecting information by allocating limited resources of processing (Anderson, 2009). The complex process of paying attention has been depicted as a continuum with different levels of attention, going from unconsciousness (total lack of awareness) to focal attention (vivid awareness) (Norris, 2004). In this section I explain why the management of attention is a key aspect of engagement, closely linked to emotion, motivation and interaction.

**Attention and multitasking.** Connecting to multiple objects at once is known as multitasking. Switching tasks can the result of external interruptions (Mark, Iqbal, Czerwinski, & Johns, 2015) or of self-interruptions such as internal decisions (Benbunan-Fich, Adler, & Mavlanova, 2011). Since multitasking depends on the management of thoughts and notifications, the variety of possibilities of attending to multiple stimuli in a short period of time is high. Users do not cope with several connections simultaneously but experience them sequentially, in a process of fast engaging and disengaging. Typically, each of the old and new connections can be explained by motivation. However, when studying the reason why a user engages into a new object, one also needs to take into account the user's emotional and attentional state prior to it.

Mark et al. (2015) studied states of attention in a work environment. In order to understand how people multitask while they perform their job tasks, the authors tracked thirty-two employees by means of different metrics. They found that, in any time of the day, the choice of a particular object was related to the one object used just a moment before. For instance, rote or routine work was followed by more Facebook or face-to-face interaction, while focused and aroused states lead to more e-mail. Mark et al. (2015) concluded that users choose some objects and create connections as 'short breaks' in their on-going tasks, breaks aimed at emotional relief (also known as emotional homeostasis) and at keeping the balance. Furthermore, even though attention is linked to the activities' degree of challenge, the availability of the other objects is an influent, possibly distracting factor.

Prior to engaging with an object, the user's attention is already susceptible to be distracted. This means that dividing the phases into "point of engagement", "engagement", "disengagement" and "re-engagement" as depicted by O'Brien and Toms (2008) would be over-simplistic. Users are potentially already unconsciously connected to a new object before it actually happens. Therefore, each connection must be explained by the context where other objects come into play, by the previous object the user has connected with, and by the previous interactions with the same object (if any). They can all be indicative of the reason why the user engages in a connection with an object.

The beginning and the end of the process of engaging with an object tend to be blurry and fragmented. However, the interactions and the elements the user identifies as emotional rewards are able to lead to higher states of attention. For example, a user can feel positive emotions after achieving the proposed challenges in a video game, which in turn would stimulate him to continue and set more difficult challenges, until perhaps reaching a Flow state. This may depend on many variables such as the user's skills or object design (i.e. challenges), which makes the Flow outcome – loss of sense of time – a very unique guarantee of a long-lasting connection.

Most computing products are used in a noisy environment with multiple objects sending notifications (e.g. e-mail or Social Media), and therefore to study their connections one has to consider multiple periods of time. Each connection is dependent on the previous connections, their interactions and sketched situations. Likewise, different connections held over time between the same user and object can be analysed as a longer connection or, in other words, as a relationship.

**Multiple connections and transitioning states of attention.** User attention states are a reflection of how connections are developed with an object and with its composition, or even in a broader context of multiple objects. Remaining in a connection or transitioning to others will depend on how the user discriminates the different stimuli provided by single or a composed object. I delimit four different states of attention - flow, rote, distracted and background - taking into account the continuum from unconsciousness (total lack of awareness) to focal attention (vivid awareness) (Norris, 2004). The four states of attention are depicted in ***Figure*** .

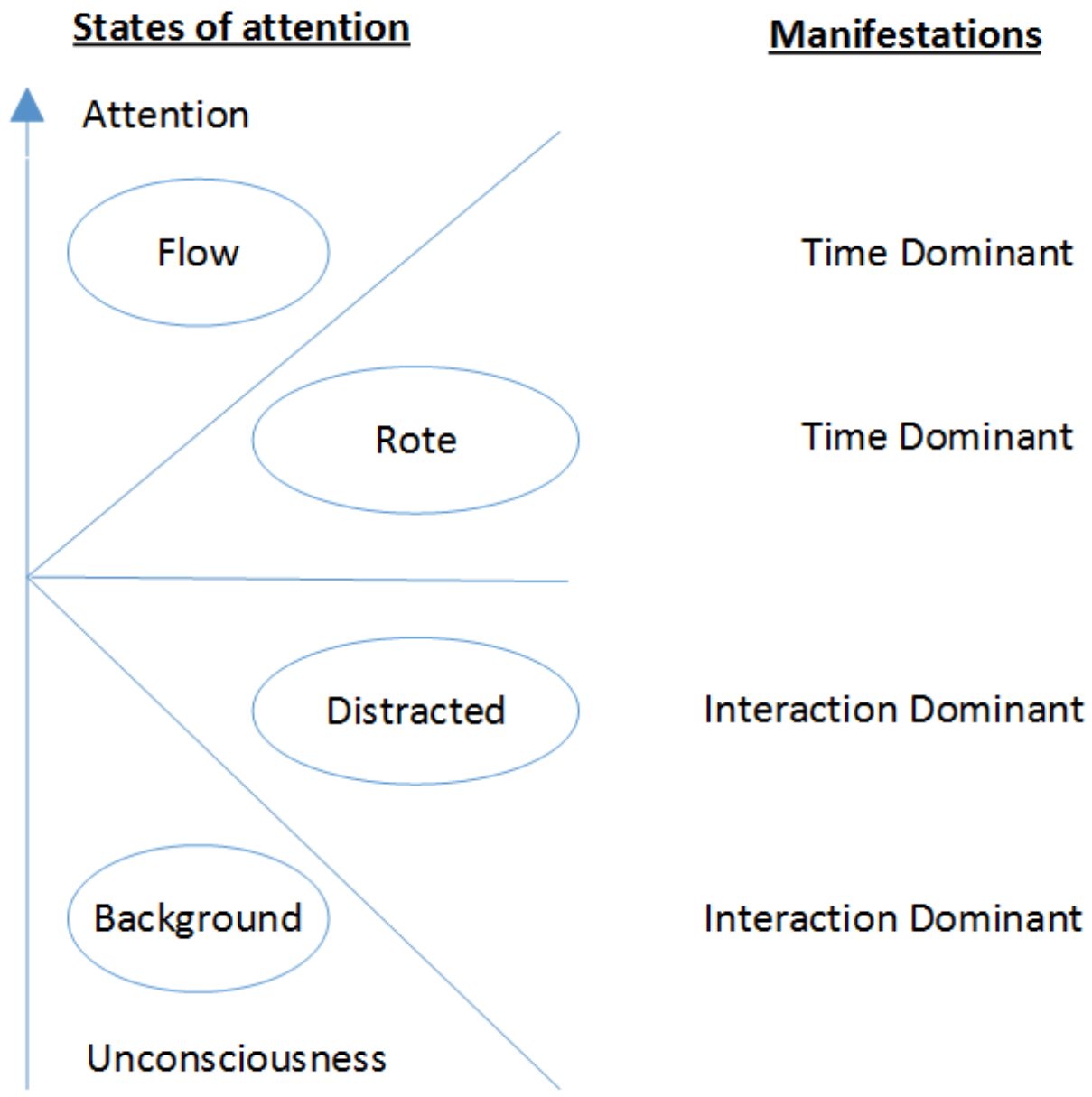

***Figure 4. States of attention and their manifestations on the user-object connection.***

Notice that in one end there is the Flow state (where the user is interacting with the objects as a whole), while in the other end there is the background state (where the user knows there is a connection opened or the possibility to start one but has not engaged in it yet). Each state of attention is a diffuse division to help understanding the experiences with computing products. Depending on the user and object aspects, as well as on the overall context described, the user can transition from one state to another, maintaining or switching between objects. Depending on the user attentional state, it pays more importance to measure one type of manifestation than another: longer connection or more interaction.

- **Flow state** manifests when the user feels a sense of direction in the experience, a connection totally excluding the other unnecessary objects (Csikszentmihalyi, 1991; Nakamura & Csikszentmihalyi, 2009). The user completely abandons himself in the connection with a very focused attention. Either the computing product or products or the user takes total control of the interaction, in a challenging progression or a repetition stimulated by the design itself. As already mentioned, Flow can be experienced in many situations, for instance while working on problem with a software tool (Mark et al. 2015). A connection involving a user in the Flow state of attention tends to last more than others.

- **Rote state** appears when the user's attention is occupied by several connections in a coordinated experience but without challenge. The rote state allows progressing in one direction towards a goal. The user can reorganise his priorities to maintain a certain focus, although there is no challenge (Mark et al. 2015). This is a common state while certain work tasks. A connection involving a user in a rote state tends to manifests first in the time spent and second in the interactions, because of the reorganization of the multiple objects to pursue the activity objective.

- **Distracted state** appears when multiple objects pop-up resulting in new connections starting while other possible connections are left for a later stage. Generally, distracted state implies pursuing several goals at a time and if instead a single goal is wanted, the user has to struggle to maintain attention on it. Between explorative and curious, the user is motivated to change the object or to explore the different inner objects which are likely to emerge from a bigger object (Marsh & Nardi, 2014). In any case, the user is externally directed by multiple objects. This state commonly manifests when surfing the Internet while working, or in a social networking site. A connection involving a user in the distracted state tends to manifest in a larger number of interactions with a shorter duration.

- **Background state** manifests either when a user is aware of a new object but chooses not to focus his attention on it, or when he remembers that an active connection has been left open and could possibly be resumed. The user can unconsciously resume a connection in background state in order to draw further information, hence interrupting an on-going activity in focus state. Smartphones and smartwatches are a clear example of devices with objects in reach, likely to stimulate the user to start a connection. With a user in a background state, the most important issue to consider is the time it takes for the user to react and interact with the object.

According to (Mark et al. 2015), user's attentional state is related to the object the user is interacting with (for instance it is hardly possible to stay focused in social networking sites). Manifestations of the connection will be as varied as the wide range of objects. Hence, depending on each object's purpose, success in terms of engagement can be either better represented by time duration, or by the number of interactions or multiple accesses. Some computing products may only be used by a user in a Flow state, while others will be often used in a distracted state.

In an object that aims at a participatory type of engagement, it will be equally useful to have connections with users during multiple periods of time with interaction (e.g. logging into Twitter several times a day for a tweet) or multiple connections with several inner objects in a period of time (e.g. making several tweets directed to different Twitter users in one single access). This is very common in objects such as Social Media or Online Communities, where there is a bigger purpose as well as different inner objects which encourage different sorts of interactions.

## 2.6 Facets of the connection

Since people live potentially attached to computing products, there is a huge interest in measuring the connections and tracking their activity. For a designer or a manufacturer, success depends on how engaging the product is. The way the engagement of an object is rated differs. Online marketing companies and some researchers have somewhat intuitively assessed the value of each manifestation in relation to the object (Lehmann et al., 2012). Namely, certain websites measure their success in terms of short visits followed by frequent returns of visitors, while others in terms of long visits. It is only relevant to compare objects with a similar composition, purpose and functionalities, or similar groups or types of users, to see how they vary in the connection's manifestations.

In studying engagement, one needs to consider the aspects related to both user and object to understand which part is more determining in keeping the connection active. Yet, due to the variety of objects, their causal relationships cannot be determined in a single way. Engagement is multi-causal. Aspects in the user (motivation, emotion, attention and cognition) and in the object (design, content, logics and composition) are the causes of the manifestations, while the latter can be taken as the consequences. I integrate them into a conceptual model suitable for analysing engagement in a user-object connection. Nevertheless, it has to be borne in mind that this model is not presented as "comprehensive model" for every object. I only included the most common aspects, additional ones should be introduced for particular object studies (e.g. challenge, interest or aesthetic pleasure) to obtain complementary insights.

In this model, I propose *four descriptive facets* to operationalize the connection manifestations, by focusing on time, interaction or a combination of them, in order to characterize the manifestations of any connection (***Figure*** ). They show that an engaging user-object connection can enhance either a *faster appeal*, a *longer duration*, a *higher interaction* or a *frequent return*. Differently put, each manifestation makes it possible to assess the success of each object. I propose broad facets to encompass all the manifestations of an active connection and previous research engagement studies. This is precisely the solution in order to integrate the two different types of engagement, the one closer to user experience in the Human-Computer Interaction tradition, and the participatory type from the Social Sciences tradition. I resort to the facets to review the specific metrics employed by the current literature to assess the level of engagement.

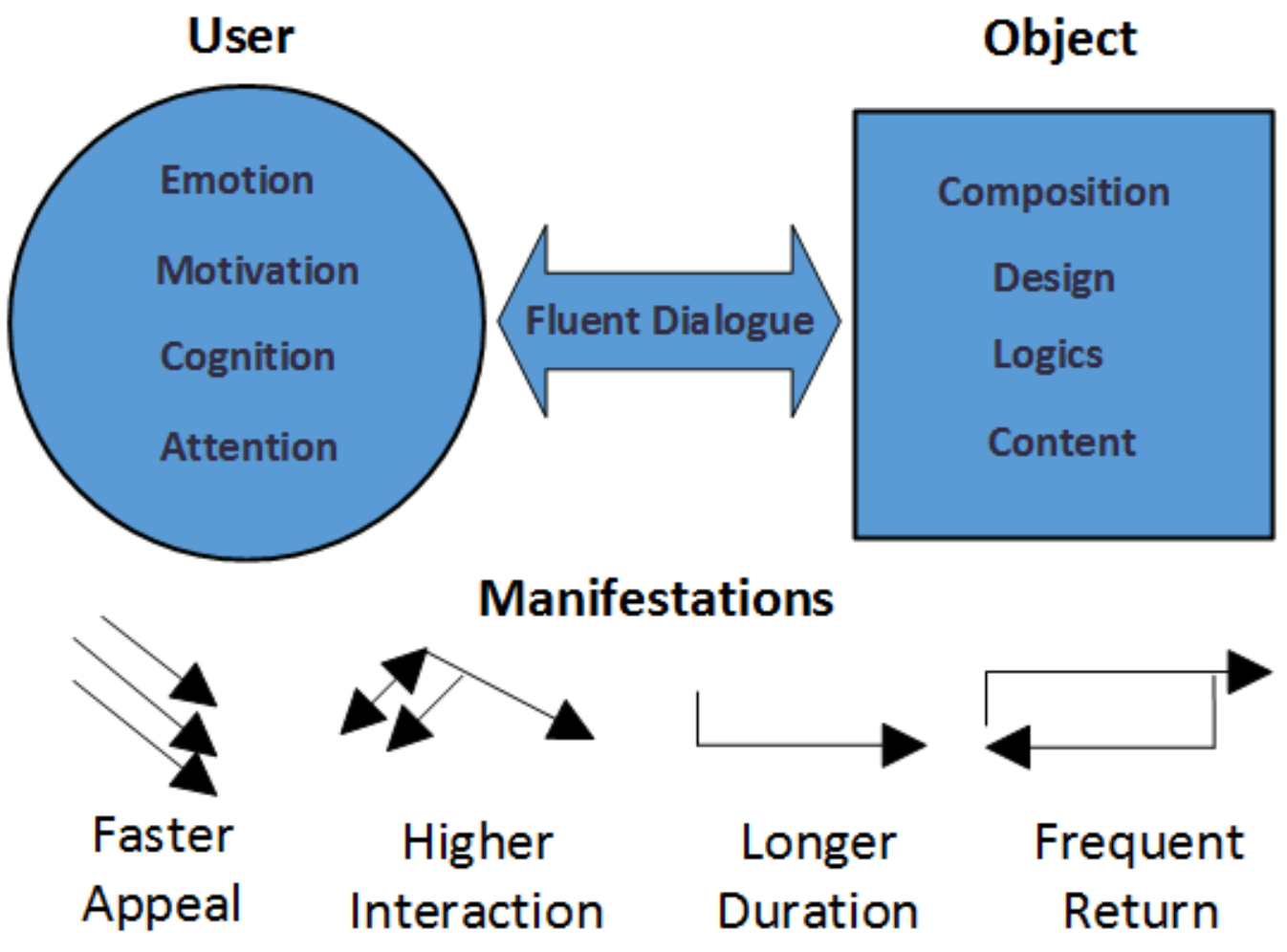

*Figure 5. Model of engagement with aspects and manifestations.*

**Faster Appeal** stresses the importance of the beginning of a connection. Since an engaging object catches and captivates the user's interest (Jacques, 1995), faster appeal refers to this initial period. Hence, measuring faster appeal is tantamount to assessing the time it takes from an initial point of the connection to a more advanced one, or to quantifying the number of connections initiated with an object. Faster appeal can be established either with an object or with its different inner objects (i.e. the time it would take to click on a picture on a social networking site, or the number of clicks a picture receives would be measures of this inner object faster appeal). For instance, faster appeal can be used to understand the first hour of video game playing (Cheung et al., 2014) or the first days as a Wikipedia editor (Panciera, Halfaker, & Terveen, 2009). In order to consider certain objects as successful, the user should not disconnect at an initial stage.

**Higher Interaction** pays attention to the number of interactions in a particular connection or in an aggregation of connections. Digital marketing mostly measures the interactions of a user with an object, but higher interaction also encompasses the notifications an object sends to the user. Hence, to measure higher interaction it is important to define whether it is within a single connection or within the sum of various connections. For instance, in online news or videos websites, a higher interaction in terms of user comments or contributions has been considered a positive sign of engagement, and is also reffered to as "participation" (Ksiazek et al., 2014). In a social networking site, Freyne et al. (2009) proposed the use of a recommendation tool in charge of sending messages aimed at increasing user interaction, which eventually led the user to make more contributions.

**Longer Duration** stresses the importance of the time spent in a connection or in an aggregation of connections. While longer duration can be measured between the engagement point and disengagement, some studies also consider the "perceived time" by the user (Arapakis et al., 2014). Time spent navigating in a website is very indicative of the type of site (Lehmann et al., 2012). For instance, in the context of video playing, Dobrian et al. (2011) proved that the video quality had an effect on playtime and such effect was more or less intense depending on the type of video (e.g. sports or a TV show). Configurations such as a lower bitrate or buffering rate decreased viewing time.

**Frequent Return** pays attention to the resumption of previous connections. Some objects may not necessarily be used during a long period of time but be continuously accessed instead. If an object is engaging it will create endurability (Lalmas et al., 2014; O'Brien & Toms, 2008). This facet is usually implemented by metrics which measure the time between sessions as well as the number of times a connection has been resumed. For instance, the intersession time (also known as 'absence time') has been measured in users consulting search websites such as Questions & Answers (Dupret & Lalmas, 2013). In a way, absence time and return rate metrics can perfectly complement the metrics from the facet faster appeal. Depending on the object, these metrics are also referred to as "loyalty", "retention" or "survival" metrics. They are especially important, for instance, in measuring a customer in an e-commerce website, or an editor in Wikipedia.

## 3. Conclusions and future research

Research on engagement with computing products sheds light on several topics of key interest in technology use. Since the late 80's, engagement with technology was based on the psychological aspects of the user. Nonetheless, the spread of different Internet applications has shaken the way and the contexts in which people use technology, either to play games, to learn or to buy any product, and consequently require to revise the current models to study engagement. Because a different use of the term, rooted in the Social Sciences, implies a participatory sense which is now indispensable to understand these social and objects.

In order to conciliate these various meanings, this paper proposed a working definition with engagement as the quality which ensures a user-object connection stays active. Hence, an engaged behaviour can be either the user absorbed or participating frenetically, but in both cases, it guarantees the connection remains active. Each connection may manifest itself in a different way (longer or shorter duration, and more or less interaction). These manifestations are measurable and can be explained by studying each part of the connection.

Hence, the proposed engagement model takes the connection as the unit of analysis. This view evolves from a user-centred perspective in studying HCI, which has been dominant first, since usability studies appeared, focusing on the object properties which enable task efficiency, and second, with user experience aiming at explaining the user's range of emotions and needs in relation to a product or a service. As said, the user-centred perspective is useful for designing as it helps in understanding aspects of cognition and needs, but assumes most of theories assume that the object is passive. This is not what happens nowadays or will happen in the future.

Engagement integrates both perspectives and considers the computing product and the user are intertwined in a connection by and for different reasons. This is because the study of engagement needs to go beyond motivation or a user-centred perspective, since objects present an active reality. I advocate that this paradigm shift will take more relevance when objects become more complex, in terms of design, both in the audio-visuals and the behaviour encoded in advanced Artificial Intelligence algorithms.

I discussed the role of several aspects from the user and from the object and their influence on the connection. In the first, psychological aspects like motivation, emotion and cognition. In the second, design, either by providing rapid feedback, usability, and by anticipating interaction. I explained how the user's attention and its different states proved to be the key factors to understand how focus is related to the type of object in terms of composition and purpose. Depending on the object's composition, purpose and the user's attention, the connection will manifest towards longer time or a higher interaction. All in all, these aspects were unified in a model, along four facets to explore the manifestations of the connection.

For some authors, engagement has been considered a *science* (Attfield et al., 2011), due to the extensive interdisciplinary literature and the increasing complexity of measuring it. The conceptual model presented in this paper is expected to help stimulate research on both direction and impetus. Once a clear definition and model are set, the real challenge lies on measurement. Studying engagement may lead to improve any computing product design, whose use can be tracked and support specific changes in an iterative process. Technological progress is taking place at a great pace, and some exciting avenues for future research into the connections between users and computers lie ahead.

**Acknowledgements**
I want to thank Mari-Carmen Marcos and David Laniado for their advice. Universitat Pompeu Fabra and research center Eurecat.